\documentclass[a4paper,11pt]{article}

\usepackage{amsmath,amssymb}     
\usepackage{color}
\usepackage{graphicx}
\usepackage{subfigure}
\usepackage{cite}                
\usepackage{hyperref}            
\usepackage{multirow,makecell}   
\usepackage{authblk}

\numberwithin{equation}{section}   

\def \be {\begin{equation}}
\def \ee {\end{equation}}
\def \ba {\begin{array}}
\def \ea {\end{array}}
\def \bea{\begin{eqnarray}}
\def \eea{\end{eqnarray}}
\def \nn {\nonumber}

\def \a {\alpha}

\def \d {\delta}
\def \D {\Delta}

\def \m {\mu}

\def \k {\kappa}
\def \l {\lambda}

\def \s {\sigma}

\def \o {\omega}
\def \O {\Omega}
\def \th {\theta}

\def \Th {\Theta}
\def \t {\tau}

\def \p {\partial}

\def \td {\tilde}

\def \inf {\infty}

\begin{document}

\title{\textbf{Semiclassical spectrum for BMN string in $Sch_5\times S^5$}}

\newcommand\email[1]{\thanks{\href{mailto:#1}{\nolinkurl{#1}}}}
\makeatletter
\makeatother
\renewcommand\Authfont{}
\renewcommand\Affilfont{\small}
\setlength{\affilsep}{2em}

\author[a,b]{Hao Ouyang\email{ouyangh@ihep.ac.cn}\,}

\affil[a]{\,Institute of High Energy Physics and Theoretical Physics Center for Science Facilities, Chinese Academy of Sciences, 19B Yuquan Road, Beijing 100049, P.~R.~China}
\affil[b]{\,School of Physical Sciences, University of Chinese Academy of Sciences, 19A Yuquan Road Beijing 100049, P.~R.~China}

\date{}
\maketitle



\begin{abstract}
We investigate the algebraic curve for string in $Sch_5\times S^5$. We compute the semiclassical spectrum for BMN string in $Sch_5\times S^5$ from the algebraic curve. We compare our results with the anomalous dimensions in $sl(2)$ sector of the null dipole deformation of $\mathcal{N} = 4$ super Yang-Mills theory.
\end{abstract}



\section{Introduction}

Spectrum of superstrings in $AdS_5\times S^5$ is related to the spectrum of scaling dimensions in planar $\mathcal{N} = 4$ supersymmetric Yang-Mills theory via the AdS/CFT duality \cite{Maldacena:1997re,Gubser:1998bc,Witten:1998qj}.
Integrability on both sides of the duality helps us dramatically finding and understanding the AdS/CFT spectrum (For a big review, see\cite{Beisert:2010jr}).
In the $\mathcal{N} = 4$ supersymmetric Yang-Mills theory, the planar anomalous dimension matrix of infinitely long composite operators corresponds to Hamiltonian of integrable spin chain\cite{Minahan:2002ve}. This implies that the spectrum can be solved efficiently by the Bethe ansatz.

On the string side, classical integrability of superstrings in $AdS_5\times S^5$ follows from the existence of an infinite number of conserved charges \cite{Bena:2003wd} generated by the monodromy matrix of the Lax connection.
Algebraic curve for classical solution of superstring in $AdS_5\times S^5$  \cite{Kazakov:2004qf,Kazakov:2004nh,Beisert:2004ag,SchaferNameki:2004ik,Beisert:2005bm} can be obtained from the Lax connection. It plays an important role in studying the semiclassical strings in $AdS_5\times S^5$.

In recent years, much attention has been enjoyed by the integrable deformations of AdS/CFT.
One intriguing example is  the Schr\"odinger spacetime \cite{Herzog:2008wg,Maldacena:2008wh,Adams:2008wt}. Schr\"odinger spacetime can be obtained from AdS background  by an appropriate TsT (T-duality-shift-T-duality) transformation \cite{Lunin:2005jy} or null Melvin twist and has been shown to be classically integrable \cite{Orlando:2010yh,Kawaguchi:2011wt,Kawaguchi:2013lba}.
String theories in Schr\"odinger spacetime is  dual to null dipole deformed field theories \cite{Alishahiha:2003ru} (see also
\cite{Bergman:2000cw,Dasgupta:2000ry,Bergman:2001rw}).
It is interesting to study the spectrum on both sides of the Schr\"odinger/dipole CFT duality with the methods of integrability.

Integrability in null dipole deformed $\mathcal{N} = 4$ super Yang-Mills was discussed in detail in \cite{Guica:2017mtd}.
The dipole deformation can be described as a Jordan cell  Drinfeld-Reshetikhin twist \cite{Drinfeld:1989st,Reshetikhin:1990ep} in the spin chain picture.
The traditional Bethe ansatz is inapplicable due to the absence of a vacuum state. One-loop spectrum of the nontrivial twisted $sl(2)$ sector was instead obtained from the Baxter equation. In the large $J$ limit, the anomalous dimension of the ground state perfectly matches the classical energy of the BMN string at order $J^{-1}$.

The purpose of this paper is to study the Schr\"odinger/dipole CFT duality by comparing  semiclassical spectrum around classical string solutions to anomalous dimension of operators in the $sl(2)$ sector at order $J^{-2}$ in the large $J$
limit.
 One reason to study the order $J^{-2}$ terms is that in the well studied AdS$_5$/CFT$_4$ correspondence, the gauge theory and string results match at order $J^{-2}$ in the BMN limit \cite{Berenstein:2002jq}. One can expect that the null dipole deformation preserves this matching.
 Another reason is that at order $J^{-2}$ we should consider one-loop quantum string theory corrections to the string energy, while the previous test at order $J^{-1}$ involve purely bosonic classical string energies.
We compute fluctuation energies of the excitations and the one-loop shift of the ground sate energy from algebraic curve.
We show that semiclassical spectrum around the BMN string solution perfectly matches the spin chain prediction.

This paper is organized as follows. In section 2 we discuss the $Sch_5\times S^5$ background and TsT transformation in detail. We discuss the algebraic curve for strings in this background and obtain the quasi-momenta for the BMN string. In section 3, we review the algebraic curve method for computing the  fluctuation energies around classical string solutions. Then
we compute the semiclassical spectrum for the BMN strings. In section 4, we compare  string theory results obtained in section 3 with the 1-loop spectrum in the $sl(2) $ sector of the null dipole deformation of $\mathcal{N} = 4$ super Yang-Mills theory.

\section{Algebraic curve for strings in $Sch_5\times S^5$}

\subsection{$Sch_5\times S^5$ from TsT transformation}
Schr\"odinger spacetime can be constructed by applying a TsT transformations to the AdS background \cite{Alishahiha:2003ru,Herzog:2008wg,Maldacena:2008wh,Adams:2008wt}. In this paper we are interested in a particular case of  $Sch_5\times S^5$ obtained by
acting a TsT transformation on $AdS_5\times S^5$
\footnote{A more general class of Schr\"odinger deformations of $AdS_5\times X_5$ has been studied in \cite{Bobev:2009mw}.}.
 We begin with the $AdS_5\times S^5$ solution of type IIB supergravity
\begin{eqnarray}
  d s^2 &=& R^2\left( \frac{-2d x^+d x^-+(d x^1)^2+(d x^2)^2+d z^2}{z^2}
+(d \psi+A)^2+ ds^2_{\mathbb{CP}^2} \right),\label{ads5s5} \\
  ds^2_{\mathbb{CP}^2} &=&d \phi_1^2+\frac{1}{4} \sin^2 \phi_1^2
(\cos \phi_1^2(d\th_1+\cos\phi_2d\th_2)^2+d\phi_2^2+\sin^2 \phi_2 d\th_2^2), \\
  A &=& \frac{1}{2}\sin^2\phi_1( d \th_1+\cos \phi_2 d \th_2).
\end{eqnarray}
The five-form field strength is given by
\begin{equation}
 F_5=4 R^4(-\frac{1}{z^5} dx^+ \wedge d x^-\wedge d x^1\wedge d x^2\wedge d z+\mathrm{vol}(S^5)).
\end{equation}
We perform a TsT transformation to this geometry. We make a first T-duality along $\psi$, followed a shift
$x^- \rightarrow x^- - \mu \psi$, and then apply a second T-duality along  $\psi$ coordinate. After this TsT transformation, the solution reads
\begin{eqnarray}
  d s^2 &=&R^2 \left(-\mu^2\frac{(dx^+)^2}{z^4}+ \frac{-2d x^+d \tilde{x}^-+(d x^1)^2+(d x^2)^2+d z^2}{z^2}
+(d \tilde{\psi}+A)^2+ ds^2_{\mathbb{CP}^2} \right),\label{ads5s5} \\
  B_2 &=&\m R^2 (d\tilde{\psi}+A)\wedge\frac{d x^+}{z^2}.
 \end{eqnarray}
The five-form $F_5$ is invariant under the transformation. The TsT translation preserves the symmetries that commute with  $\hat J$ and $\hat P_-$. The symmetries of $SO(4,2)$ that commute with $\hat P_-$ generate the Schr\"odinger group.
The Schr\"odinger group contains Galilean group as a subgroup and has two more generators  corresponding to a
non-relativistic scale transformation $\hat{D}+\hat{M}_{+-}$ and a  special conformal transformation $\hat{K}_-$.
The energy of the string is defined as the global charge associated with the symmetry $(\hat{P}_++\hat{K}_-)/\sqrt{2}$ which is related to the non-relativistic scale transformation $\hat{D}+\hat{M}_{+-}$ by a  similarity transformation.
Holography enable one to compute the non-relativistic conformal dimensions of operators at strong coupling as the energies of strings.

 The  relations between the original and dual coordinates are
\begin{eqnarray}
  d \psi &=& d \tilde{\psi} + *\mu\frac{d x^+}{z^2},\\
  d {x}^- &=& d \tilde{x}^- + *\mu(d\tilde\psi+ *\mu\frac{d x^+}{z^2}+A).
\end{eqnarray}
Here we make a slight abuse of notation that we use the same symbols for forms on the target space and their pull-back to the worldsheet.

We consider the closed strings on the deformed background. The dual coordinates satisfy  periodic boundary conditions. The original coordinates have the following twisted boundary conditions
\begin{align}
x^-(2\pi)-x^-(0)&=L J,\\
\psi(2\pi)-\psi(0)&=2\pi m_1-L P_-,~~~m_1\in\mathbb{Z},
\end{align}
where $L=2\pi \mu/ \sqrt{\l}$ is the deformation parameter in the dual field theory and
\begin{eqnarray}
J&=&\frac{\sqrt{\l}}{2 \pi}\int_0^{2\pi}d\s ({\p}_\t\psi+A_\t),\\
P_-&=&\frac{\sqrt{\l}}{2 \pi}\int_0^{2\pi}d\s \frac{-\partial_\tau x^+}{z^2} ,
\end{eqnarray}
are global charges associated with symmetries $\hat J$ and $\hat P_-$, and $\sqrt{\l}=R^2/\a'$ is the square root of 't Hooft coupling.

\subsection{Flat connection}
Integrability  of the string sigma model is preserved by TsT transformation, so strings in Schr\"odinger spacetime is integrable \cite{Orlando:2010yh,Kawaguchi:2011wt,Kawaguchi:2013lba} (see also \cite{Klimcik:2008eq,Klimcik:2002zj,Delduc:2013qra,Kawaguchi:2014qwa,Matsumoto:2014ubv,Matsumoto:2014gwa,
Matsumoto:2015uja,vanTongeren:2015soa,vanTongeren:2015uha,Osten:2016dvf,Hoare:2016wsk,
vanTongeren:2016eeb,Hoare:2016wca,Araujo:2017jkb,Araujo:2017jap}).

We now construct the Lax connection for IIB superstring  in $Sch_5\times S^5$. The type IIB superstring in $AdS_5 \times S^5$ can be described by the a sigma-model in supercoset  space of the super group $SU(2,2|4)$  over $SO(4,1)\times SO(5)$ \cite{Metsaev:1998it}. To describe strings in Poincar\'e coordinates, we
choose the coset representative as
\begin{equation}
g(x^\mu,z,\psi,\th_i,\phi_i,\Th)=B(x^\mu,z,\psi,\th_i,\phi_i)e^{F(\Theta)},
\end{equation}
with
\begin{equation}
B(x^\mu,z,\psi,\th_i,\phi_i)=e^{ix^+\hat{P}_++ix^-\hat{P}_-+ix^1\hat{P}_1+ix^2 \hat{P}_2}e^{-i\log (z)\hat{ D}}e^{i\psi \hat{J}}B_1(\th_i,\phi_i),
\end{equation}
and $\Theta$ represents the fermionic coordinates. We use a matrix representation such that
\begin{equation}
\begin{split}
&\hat{P}_-=
\left(
\begin{array}{c|c}
\begin{array}{cccc}
 0 & 0 & 0 & \sqrt{2} \\
 0 & 0 & 0 & 0 \\
 0 & 0 & 0 & 0 \\
 0 & 0 & 0 & 0 \\
\end{array}
 & 0_{4\times 4} \\
\hline
 0_{4\times 4} & 0_{4\times 4} \\
\end{array}
\right),~~~\hat{P}_+=
\left(
\begin{array}{c|c}
\begin{array}{cccc}
 0 & 0 & 0 & 0\\
 0 & 0 & \sqrt{2} & 0 \\
 0 & 0 & 0 & 0 \\
 0 & 0 & 0 & 0 \\
\end{array}
 & 0_{4\times 4} \\
\hline
 0_{4\times 4} & 0_{4\times 4} \\
\end{array}
\right),\\
&\hat{D}=\frac{i}{2}\mathrm{diag}(1,1,-1,-1|0,0,0,0),~~~
\hat{J}=\frac{1}{2}\mathrm{diag}(0,0,0,0|1,1,1,-3).
\end{split}
\end{equation}

The current associated with $g$ can be defined as
\begin{equation}
\mathcal{J}=-g^{-1}dg=e^F\mathcal{J}_Be^{-F}-e^Fde^{-F}=\mathcal{J}_B+\mathcal{J}_F,
\end{equation}
where
\begin{eqnarray}
&&\mathcal{J}_B=-B^{-1}dB=-\frac{i}{z}dx^\m \hat{P}_\m+\frac{i}{z}dz \hat{D}
-i d\psi B_1^{-1} \hat{J} B_1- B_1^{-1} d B_1 \\
&&\mathcal{J}_F=\frac{\sinh \mathcal{M}}{\mathcal{M}}\nabla F+2[F,\frac{\sinh^2 \mathcal{M}/2}{\mathcal{M}^2}\nabla F],
\end{eqnarray}
with
\begin{equation}
\nabla\cdot:=d\cdot+[\mathcal{J}_B,\cdot],~~~ \mathcal{M}^2\cdot:=[F,[F,\cdot]].
\end{equation}

The  Lie superalgebra $su(2, 2|4)$ has a $\mathbb{Z}_4$ grading structure associated with a $\mathbb{Z}_4$ automorphism $\O$.
The  automorphism $\Omega$ in this matrix representation is defined by
\begin{equation}
 \O(M)=-K M^{st} K^{-1},~~~K=
 \left(
\begin{array}{cccc|cccc}
 0 & -1 & 0 & 0 & 0 & 0 & 0 & 0 \\
 1 & 0 & 0 & 0 & 0 & 0 & 0 & 0 \\
 0 & 0 & 0 & 1 & 0 & 0 & 0 & 0 \\
 0 & 0 & -1 & 0 & 0 & 0 & 0 & 0 \\ \hline
 0 & 0 & 0 & 0 & 0 & -1 & 0 & 0 \\
 0 & 0 & 0 & 0 & 1 & 0 & 0 & 0 \\
 0 & 0 & 0 & 0 & 0 & 0 & 0 & -1 \\
 0 & 0 & 0 & 0 & 0 & 0 & 1 & 0 \\
\end{array}
\right)
\end{equation}
We can decompose the current $\mathcal{J}$ into
\begin{equation}
 \mathcal{J}=\mathcal{J}^{(0)}+\mathcal{J}^{(1)}+\mathcal{J}^{(2)}+\mathcal{J}^{(3)},
\end{equation}
where $\O(\mathcal{J}^{(i)})=i^{n} \mathcal{J}$. Then the equation of motion of string in $AdS_5 \times S^5$ is  equivalent to the conservation of the Noether current
\begin{equation}
d*k=0, ~~~k=g(\mathcal{J}^{(2)}-\frac{1}{2}\mathcal{J}^{(1)}+\frac{1}{2}\mathcal{J}^{(3)})g^{-1}.
\end{equation}
 Lax connection $\mathcal{L}(x)$ for the superstring in $AdS_5 \times S^5$ has been derived in \cite{Bena:2003wd}
\begin{equation}
\mathcal{L}=\mathcal{J}^{(0)}+\frac{x^2+1}{x^2-1}\mathcal{J}^{(2)}-\frac{2x}{(x^2-1)}*\mathcal{J}^{(2)}
+\sqrt{\frac{x+1}{x-1}}\mathcal{J}^{(1)}+\sqrt{\frac{x-1}{x+1}}\mathcal{J}^{(3)}.
\end{equation}
If the current satisfies the equation of motion, the Lax connection is flat
\begin{equation}
  d \mathcal{L}-\mathcal{L}\wedge \mathcal{L}=0.
\end{equation}
Using this flat connection, one can define the monodromy matrix
\begin{equation}
T(x)=\mathrm{P}\exp(\int_0^{2\pi}d\s \mathcal{L}_{\s}(x)).
\end{equation}
The eigenvalues of $T(x)$ do not depend on $\t$ and generate an infinite number of conserved quantities. The current components $\mathcal{J}_{\alpha}$ do not have an explicit dependence on $x^-$ and $\psi$. Then the Lax connection $\mathcal{L}$ in the undeformed case can be used to derive a Lax connection and thus quasi-momenta for strings in
 $Sch_5\times S^5$ background.

The quasi-momenta $p_i(x)$ are functions defined from the eigenvalues
\begin{equation*}
  \{e^{\hat p_1},e^{\hat p_2},e^{\hat p_3},e^{\hat p_4}|e^{\tilde p_1},e^{\tilde p_2},e^{\tilde p_3},e^{\tilde p_4}\}
\end{equation*}
of the monodromy matrix $T(x)$. They are generating functions of conserved physical quantities. For instance, we can read the conserved global charges from the behavior at large $x$.
Large $x$ asymptotic properties of the quasi-momenta for strings with twisted boundary condition are more complex than those for close strings. Below we analysis the asymptotic behavior of the quasi-momenta.

At $x\rightarrow \infty$, the expansion of the Lax connection is
\begin{equation}
\mathcal{L}=-g^{-1}d g-\frac{2}{ x}g^{-1}*k g+{O}(\frac{1}{x^2}).
\end{equation}
Expanding the monodromy matrix, we get
\begin{equation}
  \begin{split}
    \tilde{T}\equiv& g(\t,0)T g(\t,0)^{-1}\\
=&g(\t,0)g(\t,2\pi)^{-1}
\left(1-\frac{2}{ x}\int_0^{2 \pi} d \s k_\t+{O}(\frac{1}{x^2})\right)\\
=&e^{i L(P_-\hat{J}-J\hat{P}_-)}
\left(1-\frac{2}{ x}\int_0^{2 \pi} d \s k_\t+{O}(\frac{1}{x^2})\right)
  \end{split}
\end{equation}

In our representation, it takes the form
\begin{equation}
\tilde{T}=
\left(
\begin{array}{c|c}
\begin{array}{cccc}
 0 & 0 & 0 & - i \sqrt{2}L J \\
 0 & 0 & 0 & 0 \\
 0 & 0 & 0 & 0 \\
 0 & 0 & 0 & 0 \\
\end{array}
 & 0_{4\times 4} \\
\hline
 0_{4\times 4} &
 \exp(\frac{i}{2} LP_-\mathrm{diag}\left(1,1,1,-3)\right) \\
\end{array}
\right)+\frac{1}{\sqrt{\l}x} Q+O(\frac{1}{x^2}),
\end{equation}
where $Q$ satisfies $Q_{\hat{4}\hat{1}}=-2\pi i \sqrt{2}P_-$. Behavior of the quasi-momenta  at $x\rightarrow \infty$ reads
\begin{eqnarray}
&&\left(\begin{array}{c}
\hat p_1 \\
\hat p_2 \\
\hat p_3 \\
\hat p_4 \\
\hline
\tilde p_1 \\
\tilde p_2 \\
\tilde p_3 \\
\tilde p_4 \\
\end{array}\right)=
\left(\begin{array}{c}
2\sqrt{\frac{\pi L {J} {P}_{-}}{\sqrt{\l} x} }\\
0\\
0\\
-2\sqrt{\frac{\pi L {J} {P}_{-}}{\sqrt{\l} x}} \\
\hline
\frac{1}{2}L{{P}_{-}} \\
\frac{1}{2}L{{P}_{-}} \\
\frac{1}{2}L{{P}_{-}} \\
-\frac{3}{2}L{{P}_{-}} \\
\end{array}\right)
+
\frac{2 \pi}{\sqrt{\l} x}\left(\begin{array}{c}
O(1)\\
\Delta+S\\
-\Delta+S\\
O(1)\\
\hline
+J_1+J_2-J_3 \\
+J_1-J_2+J_3 \\
-J_1+J_2+J_3 \\
-J_1-J_2-J_3 \\
\end{array}\right)
+O(x^{-2}),
\end{eqnarray}
where $S$ and $J_i$ are spins of the string.
The quasi-momenta  $\hat p_1$ and $\hat p_4$ are connected by a square root cut which has a branch point  at infinity.  The quasi-momenta satisfy
\begin{equation}\label{levelmatching}
\oint_{\inf} dx( \tilde p_1^2+ \tilde p_2^2+\tilde p_3^2+ \tilde p_4^2
- \hat p_1^2- \hat p_2^2- \hat p_3^2- \hat p_4^2)=0,
\end{equation}
so we still have the  constraint between length and filling fractions (see \cite{Beisert:2004ag}\cite{Beisert:2005bm}).

\subsection{BMN string}

We now exemplify the discussion above with BMN string solution in $Sch_5\times S^5$  presented  in \cite{Guica:2017mtd}
\begin{equation}
\begin{split}
&\phi_i=\th_i=x^2=x^1=0,~~~\psi=\mu  m \sigma + \omega\tau ,\\
&x^+=\frac{ \tan (\kappa  \tau )}{\sqrt{2}},~~~ x^-=\mu  \omega \sigma +\frac{\kappa  \tan (\kappa  \tau )}{2 m},\\
&z=\sqrt{\frac{\kappa}{\sqrt{2}m}} \sec (\kappa  \tau ).
\end{split}
\end{equation}
Virasoro constraint gives
\begin{equation}\label{vBMN}
\mu^2 m^2 -\k^2 +\o^2=0.
\end{equation}
The conserved global charges are
\begin{equation}
\Delta=\sqrt{\l}\k,~~~P_{-}\equiv -M=-\sqrt{\l} m ,~~~J=\sqrt{\l}\o,
\end{equation}
where we denote $\Delta$ as the global energy of the string associated with the non-relativistic scale transformation $\hat{D}+\hat{M}_{+-}$. Then the classical energy of the BMN string is given by
\begin{equation}
\Delta=\sqrt{J^2+\mu^2 M^2}.
\end{equation}

The quasi-momenta of the BMN string are
\begin{equation}
\begin{split}
&\hat p_1=-\hat p_4=\frac{2\pi \k x  \sqrt{1- x\sin 2 \gamma} }{x^2-1},\\
&\hat p_2=-\hat p_3=\frac{2\pi \k  \sqrt{x}  \sqrt{x- \sin 2 \gamma } }{x^2-1},\\
&\td p_1=\frac{ \pi \o  (2 x-(1+x^2)\tan \gamma )}{x^2-1},\\
&\td p_2=\frac{ \pi \o  (2 x-(1+x^2)\tan \gamma )}{x^2-1},\\
&\td p_3=-\frac{\pi \o  (2 x+(-3+x^2)\tan \gamma )}{x^2-1},\\
&\td p_4=-\frac{\pi \o  (2 x+(1-3x^2)\tan \gamma)}{x^2-1},
\end{split}
\end{equation}
with
\begin{equation}
\sin 2 \gamma=\frac{4 \pi  \sqrt{\lambda }J L M}{4 \pi ^2 J^2+\lambda  L^2 M^2},~~~\tan \gamma=\frac{\sqrt{\lambda } L M}{2 \pi  J}.
\end{equation}
We find an expected square root cut $[\csc 2\gamma,\infty]$ connecting $\hat p_1$ and $\hat p_4$.

\section{Semi-classical quantization of the BMN string }
A powerful method for computing the semiclassical spectrum around string solutions is proposed in \cite{Gromov:2007aq}. Here we begin with a review of this method for the reader's convenience.

The semiclassical spectrum around the BMN string is given by
\begin{equation}
\Delta=\Delta_{\mathrm{cl}}+\Delta_{\mathrm{1-loop}}+\d \Delta,
\end{equation}
where $\Delta_{\mathrm{cl}}+\Delta_{\mathrm{1-loop}}$ is the ground state energy, and $\d \Delta$ is the energy of the excitations. To compute $\d \Delta$, we add  a perturbation $\d p(x)$ to $p(x)$  associated with the classical solution. The  perturbation $\d p(x)$ has a single pole at $x_n$. The position $x_n$ is determined by
\begin{equation}\label{xn}
p_i(x^{ij}_n)-p_j(x^{ij}_n)=2 \pi n.
\end{equation}
The residues at  the poles are
\begin{equation}
\mathrm{res}_{x=x^{ij}_n}~ \d \hat p_k= (\d_{ik}-\d_{jk})\a (x^{ij}_n)N^{ij}_n,~~~
\mathrm{res}_{x=x^{ij}_n}~ \d \tilde p_k= (\d_{jk}-\d_{ik})\a (x^{ij}_n)N^{ij}_n,
\end{equation}
where $i<j$ and $N_n^{ij}$ is the excitation number for excitation with polarizations $(ij)$ and mode number $n$ and the function
$\alpha(x)$ is defined as
\begin{equation}
\alpha(x)=\frac{4\pi}{\sqrt{\l}}\frac{x^2}{x^2-1}.
\end{equation}
 the residues at $x=\pm 1$ are synchronized as
\begin{equation}
\delta\{\hat p_1,\hat p_2,\hat p_3,\hat p_4|\tilde p_1,\tilde p_2,\tilde p_3,\tilde p_4\}=
\frac{1}{x\pm 1}\delta\{\alpha_{\pm},\alpha_{\pm},\beta_{\pm},\beta_{\pm}|\alpha_{\pm},\alpha_{\pm},\beta_{\pm},\beta_{\pm}\}
+...
\end{equation}

For each cut $\mathcal{C}^{ij}$ connecting $p_i(x)$ and $p_j(x)$ the  perturbation $\d p(x)$ satisfies
\begin{equation}\label{Acycle}
\d p^+_i-\d p^-_j=0,~~~x\in \mathcal{C}^{ij}.
\end{equation}
where the superscript $\pm$ denotes above and below the cut.

The poles result in an energy shift
\begin{equation}\label{energyshift}
\delta\Delta=\sum_{n,ij} N^{i j}_{n}\Omega^{ij}(x^{ij}_{n}),
\end{equation}
where $\Omega^{ij}$ are the off-shell fluctuation energies.

The asymptotic behaviour of the perturbation $\d p(x)$ is
\begin{eqnarray}
&&\delta\hat p_1=O(x^{-1/2}), ~~~\delta\hat p_4=O(x^{-1/2}),\\
&&\left(\begin{array}{c}
\delta\hat p_2 \\
\delta\hat p_3 \\
\hline
\delta\tilde p_1 \\
\delta\tilde p_2 \\
\delta\tilde p_3 \\
\delta\tilde p_4 \\
\end{array}\right)
=\frac{4 \pi}{\sqrt{\l}x}\left(\begin{array}{c}
+\frac{1}{2}\d\Delta +N^{\hat 2 \hat 3}+N^{\hat 2 \hat 4}+ N^{\hat 2 \tilde 3}+ N^{\hat 2 \tilde 4} \\
-\frac{1}{2}\d\Delta -N^{\hat 2 \hat 3}-N^{\hat 1 \hat 3}- N^{\tilde 1 \hat 3}- N^{\tilde 2 \hat 3}\\
 \hline
-N^{\tilde 1 \tilde 3}- N^{\tilde 1 \tilde 4}-N^{\tilde 1 \hat 3}- N^{\tilde 1 \hat 4}\\
-N^{\tilde 2 \tilde 3}- N^{\tilde 2 \tilde 4}-N^{\tilde 2 \hat 3}- N^{\tilde 2 \hat 4}\\
+N^{\tilde 1 \tilde 3}+ N^{\tilde 2 \tilde 3}+N^{\hat  1\tilde 3}+ N^{\hat 2 \tilde 3}\\
+N^{\tilde 1 \tilde 4}+ N^{\tilde 2 \tilde 4}+N^{\hat 1 \tilde 4}+ N^{\hat 2 \tilde 4}\\
\end{array}\right)+\mathcal{O}(x^{-2}).
\end{eqnarray}
The order $O(x^{1/2})$ terms in the $\d \hat p_1$ and $\d \hat p_4$ are determined by the  constraint (\ref{levelmatching}). Since the classical solution already obey the constraint (\ref{levelmatching}), one can show these filling fractions $N^{ij}_{n}$ satisfy
\begin{equation}
\sum_{n,ij} n N^{ij}_{n} =0.
\end{equation}
 using Riemann bilinear identity (see eqs. 3.38 and 3.44 in \cite{Beisert:2005bm}).

We now determine the most general form of the perturbation $\d p(x)$ for the BMN string. When small poles are added to $\d \hat p_2(x)$ with a square root cut, the branch point will be slightly displaced so $\d \hat p_2(x)$ behave like $\partial_{x_0}\sqrt{x-x_0}\sim1/\sqrt{x-x_0}$ near the branch point $x_0=\sin2\gamma$. The most general form for $\d \hat p_2(x)$ is $f(x)+g(x)/K(x)$ where $f$ and $g$ are rational functions of $x$ and $K(x)=\sqrt{x}\sqrt{x-x_0 }$. From (\ref{Acycle}) and inversion symmetry we get
\begin{equation}
\left(\begin{array}{c}
\delta\hat p_1 \\
\delta\hat p_2 \\
\delta\hat p_3 \\
\delta\hat p_4 \\
\end{array}\right)
=\left(\begin{array}{c}
-f(1/x)-\frac{g(1/x)}{K(1/x)}\\
f(x)+\frac{g(x)}{K(x)}\\
f(x)-\frac{g(x)}{K(x)} \\
-f(1/x)+\frac{g(1/x)}{K(1/x)} \\
\end{array}\right).
\end{equation}

One can obtained all other off-shell fluctuation energies  from the knowledge of $\Omega^{\tilde 2 \tilde 3}$ and $\Omega^{\hat 2 \hat 3}$ alone using the efficient method provided in \cite{Gromov:2008ec}. From inversion symmetry we have
\begin{equation}\label{inversion}
 \O^{\tilde 1 \tilde 4}(y)=-\Omega^{\tilde 2 \tilde 3}(1/y)+\Omega^{\tilde 2 \tilde 3}(0),
 ~~~ \O^{\tilde 1 \tilde 4}(y)=-\Omega^{\tilde 2 \tilde 3}(1/y)-2.
\end{equation}
the quasi-momenta for BMN string in $Sch_5 \times S^5$ are pairwise symmetric up to constant terms:
\begin{eqnarray}
 && \hat p_1 =-\hat p_4,~~~\hat p_2=-\hat p_3, \\
 && \tilde p_1=-\tilde p_4+ 2 \pi  \kappa  \sin \gamma ,
~~~\tilde p_2=-\tilde p_3 -2 \pi  \kappa  \sin \gamma .
\end{eqnarray}
So the off-shell fluctuation energies satisfies
\begin{equation}\label{relation}
  \O^{ij}(y)=\frac{1}{2}(\O^{ii'}(y)+\O^{j'j}(y)),
\end{equation}
where
\begin{equation}
  (\hat 1,\hat 2,\tilde 1, \tilde 2, \hat 3,\hat 4,\tilde 3, \tilde 4)=
    (\hat 4,\hat 3,\tilde 4, \tilde 3, \hat 2,\hat 1,\tilde 2, \tilde 1).
\end{equation}

\subsection{Fluctuation energies of excitations}

We first consider the  excitation $\hat 2 \hat 3$.
We have $\d \tilde p_i=0$, for $i=1,2,3,4$, and therefore $f(x)=0$. We take the following ansatz
\begin{equation}
g(x)=\sum_{n}\frac{\a(x^{\hat 2 \hat 3}_{n}) K(x^{\hat 2 \hat 3}_{n}) N^{\hat 2 \hat 3}_{n}}{x-x^{\hat 2 \hat 3}_{n}}+\frac{2 \pi}{\sqrt{\l}}\d \Delta+\frac{4 \pi}{\sqrt{\l}}\sum_{n} N^{\hat 2 \hat 3}_{n} .
\end{equation}
Large $x$ asymptotic of $\d \hat p_1$  gives
\begin{equation}
\d \D=\sum_{n} N^{\hat 2 \hat 3}_{n}\Omega^{\hat 2 \hat 3}(x^{\hat 2 \hat 3}_{n}) ,
\end{equation}
\begin{equation}
\Omega^{\hat 2 \hat 3}(x)=\frac{2-2x^2+2x K(x)}{x^2-1},
\end{equation}
and the level matching condition
\begin{equation}
\sum_{n} n N^{\hat 2 \hat 3}_{n} =0.
\end{equation}

We next consider the $S^3$ excitation $\tilde 2 \tilde 3$.
We start with the following ansatz
\begin{eqnarray}
&&g(x)=\frac{\a_1K(1)}{x-1}+\frac{\a_2K(-1)}{x+1}+\frac{2 \pi}{\sqrt{\l}}\d\D,\\
&&\d\tilde p_2=-\d\tilde p_3=\frac{\a_1}{x-1}+\frac{\a_2}{x+1}-\sum_{n}\frac{\a(x^{\tilde 2 \tilde 3}_{n}) N^{\tilde 2 \tilde 3}_{n}}{x-x^{\tilde 2 \tilde 3}_{n}}.
\end{eqnarray}
Large $x$ asymptotic  of $\d \tilde p_1$ and $\d \tilde p_4$ give
\begin{eqnarray}
&& \a_1+\a_2=\sum_{n}\a(x^{\tilde 2 \tilde 3}_{n}) N^{\tilde 2 \tilde 3}_{n}\frac{1}{(x^{\tilde 2 \tilde 3}_{n})^2},\\
&& \a_1-\a_2=\sum_{n}\a(x^{\tilde 2 \tilde 3}_{n}) N^{\tilde 2 \tilde 3}_{n}\frac{1}{x^{\tilde 2 \tilde 3}_{n}}.
\end{eqnarray}
Large $x$ asymptotic  of $\d \hat p_1$  give
\begin{equation}
 -\a_1K(1)+\a_2 K(-1)+\frac{2 \pi}{\sqrt{\l}}\d\D=-\a_1(\cos\gamma-\sin\gamma)-\a_2 (\cos\gamma+\sin\gamma)+\frac{2 \pi}{\sqrt{\l}}\d\D=0.
\end{equation}
The filling fractions satisfy
\begin{equation}
\sum_{n} n N^{\tilde 2 \tilde 3}_{n} =0,
\end{equation}
we can solve these equations and obtain
\begin{equation}
\d\D=\sum_{n}  N^{\tilde 2 \tilde 3}_{n}\Omega^{\tilde 2 \tilde 3}(x^{\tilde 2 \tilde 3}_{n}),
\end{equation}
where
\begin{equation}
\Omega^{\tilde 2 \tilde 3}(x)=\frac{2 \cos \gamma-2 x\sin\gamma}{x^2-1}.
\end{equation}

Finally using (\ref{inversion}) and (\ref{relation}) we find all the off-shell fluctuation energies
\begin{equation}
\begin{split}
&\Omega^{\hat 2 \hat 3}(x)=\frac{2-2x^2+2x \sqrt{x}\sqrt{x-\sin2{\gamma}}}{x^2-1},\\
&\Omega^{\hat 1 \hat 4}(x)=\frac{2\sqrt{1-x \sin2{\gamma}}}{x^2-1},\\
&\Omega^{\hat 1 \hat 3}(x)=\Omega^{\hat 2 \hat 4}(x)
=\frac{1-x^2+x \sqrt{x}\sqrt{x-\sin2{\gamma}}+\sqrt{1-x \sin2{\gamma}}}{x^2-1},\\
&\Omega^{\tilde 2 \tilde 3}(x)=\Omega^{\tilde 1 \tilde 4}(x)
=\Omega^{\tilde 1 \tilde 3}(x)
=\Omega^{\tilde 2 \tilde 4}(x)=\frac{2 \cos {\gamma}-2 x\sin{\gamma}}{x^2-1},\\
&\Omega^{\hat 2 \tilde 3}(x)=\Omega^{\tilde 2 \hat 3}(x)=
\Omega^{\hat 2 \tilde 4}(x)=\Omega^{\tilde 2 \hat 4}(x)=\frac{1+ \cos {\gamma}- x\sin{\gamma}
-x^2+x \sqrt{x}\sqrt{x-\sin2{\gamma}}}{x^2-1},\\
&\Omega^{\hat 1 \tilde 4}(x)=\Omega^{ \tilde 1\hat 4}(x)=
\Omega^{\hat 1 \tilde 3}(x)=\Omega^{ \tilde 1\hat 3}(x)=\frac{ \cos {\gamma}- x\sin{\gamma}
+ \sqrt{1-x\sin2{\gamma}}}{x^2-1}.
\end{split}
\end{equation}
We now solve the pole position $x_n$. We choose the solution $|x_n|>1$ for small $L M$ to be physical poles. We have
\begin{equation}
 \begin{split}
&x_n^{\hat 2 \hat 3}=
\frac{2}{n \sqrt{\lambda}}J+O\left(1\right),~~~ x_n^{\hat 1 \hat 4}=
\frac{2 \sqrt{L^2 M^2 \pi^{-2}+n^2}-2 L M\pi^{-1}}{\sqrt{\lambda}n^2} J+O\left(1\right),\\
&x_n^{\hat 1 \hat 3}=x_n^{\hat 2 \hat 4}=
\frac{ (2 \pi  n-L M)}{\pi  \sqrt{\lambda } n^2}J+O\left(1\right),\\
&x_n^{\tilde 2 \tilde 3}=x_n^{\tilde 1 \tilde 3}=\frac{\sqrt{\pi  J^2+\lambda  n (\pi  n-L M)}+\sqrt{\pi } J}{\sqrt{\pi } \sqrt{\lambda } n},~~~ x_n^{\tilde 1 \tilde 4}=x_n^{\tilde 2 \tilde 4}=\frac{\sqrt{\pi } \sqrt{\pi  J^2+\lambda  n (L M+\pi  n)}+\pi  J}{\sqrt{\lambda } (L M+\pi  n)},\\
&x_n^{\tilde 2 \hat 3}=x_n^{ \tilde 1 \hat 3}=\frac{\sqrt{16 \pi ^2 J^2+\lambda  (L M-4 \pi  n)^2}+4 \pi  J}{\sqrt{\lambda } (L M+4 \pi  n)},\\
&x_n^{ \hat 2\tilde 3}=\frac{\sqrt{16 \pi ^2 J^2+\lambda  (3 L M-4 \pi  n)^2}+4 \pi  J}{\sqrt{\lambda } (4 \pi  n-L M)},
~~~x_n^{ \hat 2\tilde 4}=\frac{\sqrt{16 \pi ^2 J^2+\lambda  (L M+4 \pi  n)^2}+4 \pi  J}{\sqrt{\lambda } (3 L M+4 \pi  n)},\\
&x_n^{\hat 1\tilde 4}=\frac{(L M+4 \pi  n) \left(\sqrt{16 \pi ^2 J^2+\lambda  (3 L M+4 \pi  n)^2}+4 \pi  J\right)}{\sqrt{\lambda } (3 L M+4 \pi  n)^2},\\
&x_n^{ \hat 1\tilde 3}=\frac{(4 \pi  n-3 L M) \left(\sqrt{16 \pi ^2 J^2+\lambda  (L M-4 \pi  n)^2}+4 \pi  J\right)}{\sqrt{\lambda } (L M-4 \pi  n)^2},\\
&x_n^{\tilde 2 \hat 4}=x_n^{\tilde 1 \hat 4}=\frac{(4 \pi  n-L M) \left(\sqrt{16 \pi ^2 J^2+\lambda  (L M+4 \pi  n)^2}+4 \pi  J\right)}{\sqrt{\lambda } (L M+4 \pi  n)^2}.
 \end{split}
\end{equation}
The exact expressions of $x_n^{\hat i \hat j}$  are very complex, so we only consider the leading order terms in the large $J$ expansion. When $L M \neq 0$,  a finite number of $x_{n}^{\hat 1 j}$ ($x_{n}^{i \hat 4 }$) will enter the cut connecting $\hat p_1$ and $\hat p_4$ and become $x_{n}^{\hat 4 j}$ ($x_{n}^{i \hat 1 }$).

Pluging $x_n$ into the off-shell fluctuation energies, in the large $J$ limit we get the on-shell fluctuation frequencies
\begin{equation}
  \begin{split}
&\Omega^{\hat 2 \hat 3}(x_n^{\hat 2 \hat 3})=\frac{\lambda   (\pi  n^2-L M n)}{2 \pi  J^2}+O\left(J^{-3}\right),
\\
& \Omega^{\hat 1 \hat 4}(x_n^{\hat 1 \hat 4})= \frac{\lambda}{2J^2} (\sqrt{\frac{L^2 M^2 n^2}{\pi ^2}+n^4}+\frac{LMn}{\pi})+O\left(J^{-3}\right),\\
& \Omega^{\hat 1 \hat 3}(x_n^{\hat 1 \hat 3})=\Omega^{\hat 2 \hat 4}(x_n^{\hat 2 \hat 4})=\frac{\lambda  n^2}{2 J^2}+O\left(J^{-3}\right),\\
& \Omega^{\tilde 2 \tilde 3}(x_n^{\tilde 2 \tilde 3}) =\Omega^{\tilde 1 \tilde 3}(x_n^{\tilde 1 \tilde 3})
 =\frac{\lambda   (\pi  n^2- L M n)}{2 \pi  J^2}+O\left(J^{-3}\right),\\
& \Omega^{\tilde 2 \tilde 4}(x_n^{\tilde 2 \tilde 4})
 =\Omega^{\tilde 1 \tilde 4}(x_n^{\tilde 1 \tilde 4}) =\frac{\lambda   (\pi  n^2+L M n)}{2 \pi  J^2}+O\left(J^{-3}\right),\\
&\Omega^{\tilde 2\hat 3}(x_n^{\tilde 2\hat 3})
 =\Omega^{\tilde 1\hat 3}(x_n^{\tilde 1\hat 3})
 =\frac{\lambda  \left(16 \pi ^2 n^2-8 \pi  L M n-3 L^2 M^2\right)}{32 \pi ^2 J^2}+O\left(J^{-3}\right),
\\
&\Omega^{\hat 2 \tilde 3}(x_n^{\hat 2 \tilde 3})
 =\frac{\lambda  \left(16 \pi ^2 n^2 -24 \pi  L M n +5 L^2 M^2\right)}{32 \pi ^2 J^2}+O\left(J^{-3}\right),
\\
&\Omega^{\hat 2 \tilde 4}(x_n^{\hat 2 \tilde 4}) =\frac{\lambda  \left(16 \pi ^2 n^2+8 \pi  L M n-3 L^2 M^2\right)}{32 \pi ^2 J^2}+O\left(J^{-3}\right),
\\
& \Omega^{\hat 1 \tilde 4}(x_n^{\hat 1 \tilde 4}) = \frac{\lambda  (4 \pi  n+3 L M)^2}{32 \pi ^2 J^2}+O\left(J^{-3}\right),\\
& \Omega^{\hat 1 \tilde 3}(x_n^{\hat 1 \tilde 3})= \frac{\lambda  (4 \pi  n-L M)^2}{32 \pi ^2 J^2}+O\left(J^{-3}\right),\\
&\Omega^{\tilde 2 \hat 4}(x_n^{\tilde 2 \hat 4})=\Omega^{\tilde 1\hat 4}(x_n^{\tilde 1\hat 4}) =\frac{\lambda  (4 \pi  n+L M)^2}{32 \pi ^2 J^2}+O\left(J^{-3}\right).
   \end{split}
\end{equation}
Then we obtained the energy shift $\d\Delta$  given by (\ref{energyshift}).

\subsection{One-loop shift}
The one loop shift  is equal to one half of the graded sum of all fluctuation mode frequencies. Using zeta function regularization,
we have
\begin{equation}
 \sum_{n\in \mathbb{Z}}\left((n+q)^2+p n \right)=q^2+\zeta (-2,1+q)+\zeta (-2,1-q)=0.
\end{equation}
Therefore when we compute the  one loop shift energy at order $J^{-2}$, only the contribution from $\Omega^{\hat 1 \hat 4}$  is nontrivial. Then we sum over the energies of the $sl(2)$ modes to get the one-loop shift:
\begin{eqnarray}\label{1ls}
 \Delta_{\mathrm{1-loop}}&=&\frac{1}{2}\sum_{n\in \mathbb{Z}}\Omega^{\hat 1 \hat 4}(x_{n}^{\hat 1 \hat 4}) \nn\\
&=&\sum_{n=1}^{\inf}(\sum_{k=0}^{\inf}\binom {1/2}{k}n^{2-2k}\pi^{-2k}L^{2k}M^{2k})\frac{ \lambda}{2J^2}+O\left(J^{-3}\right)\nn\\
&=&(\sum_{k=0}^{\inf}
\binom {1/2}{k}\zeta(2k-2)\pi^{-2k} L^{2k}M^{2k})\frac{\lambda }{2J^2}+O\left(J^{-3}\right)\nn\\
&=&(-\frac{\lambda L^2  M^2}{8 \pi ^2 }-\frac{\lambda L^4  M^4}{96 \pi ^2 }+\frac{\lambda L^6  M^6}{2880 \pi ^2 }+O\left(L^8M^8\right))\frac{1}{J^2}+O\left(J^{-3}\right).
\end{eqnarray}

\section{ Comparison between the string and the gauge theory results}

Comparison between the results obtained in the gauge theory and string theory is possible in the large spin regime with $J\rightarrow \infty$ and $\l/J^2$ kept fixed and small (see e.g. \cite{Berenstein:2002jq, Gubser:2002tv,Frolov:2002av,Frolov:2003qc}).
Type IIB superstring in $Sch_5 \times S_5$ is dual to null dipole deformed $\mathcal{N} = 4$ super Yang-Mills.
The $sl(2)$ sector nontrivially affected by the deformation has been studied in \cite{Guica:2017mtd}. The one-loop
spectrum of the $sl(2)$ sector can be obtained by  Baxter equation.
It is proposed in \cite{Guica:2017mtd} that the Baxter equation takes the same form as in the undeformed case
\begin{equation}
t(u) Q(u)=(u+i/2)^J Q(u+i)+(u-i/2)^J Q(u-i),
\end{equation}
and
\begin{equation}
t(u)=2 u^J+ L M J u^{J-1}+.. .
\end{equation}
The 1-loop energy is given by
\begin{equation}
\Delta^{(1)}=\frac{i \lambda}{8\pi^2}\partial_u \left.\log\frac{Q(u+i/2)}{Q(u-i/2)}\right|_{u=0}.
\end{equation}

We now solve the Baxter equation in the expansion in $LM$. At each order in $LM$, the $Q$-function is simply a polynomial. We write the ansatz
\begin{equation}
Q(u)=\sum_{k=0}^\inf p_{k+m}(u) L^k M^k,
\end{equation}
where $ p_{k+m}$ is a polynomial in $u$ of degree $k+m$. The small $LM$ expansion of $Q$ can be interpreted as a function with a finite number of zeros near the Bethe roots in the undeformed limit
and an infinite number of zeros of order $L^{-1} M^{-1}$. In the string picture the zeros of order $L^{-1} M^{-1}$   correspond to the cut connecting $\hat p_1$ and $\hat p_4$.

Substituting the above ansatz into the Baxter equation, we can determine $Q(u)$ up to multiplication by a function in $L M$.
We consider the following two solutions:
\begin{eqnarray}
&&Q_0=1-L M u+L^2 M^2 \frac{J u^2}{2 (J+1)}+O\left(L^3M^3\right),\\
&&Q_1=\left(u-u_n\right)+\frac{L M \left(-2 J u^2+2 J u_n^2+4 u_n^2+1\right)}{2
   (J+2)}\nn\\
   &&\phantom{=}+L^2 M^2 \frac{1}{2 (J+2)^3 (J+3)}(J^4 (u^3+u_n u^2-2 u_n^3)+2 J^3 (2
   u^3+3 u_n u^2-9 u_n^3-u_n)\nn\\&&
  \phantom{=} +J^2 \left(4 u^3+12 u_n u^2-60 u_n^3-11
   u_n\right)-22 J \left(4 u_n^3+u_n\right)-12 \left(4
   u_n^3+u_n\right))+O\left(L^3M^3\right),
\end{eqnarray}
with
\begin{equation}
u_n=\frac{1}{2} \cot \left(\frac{\pi  n}{J}\right).
\end{equation}
In the undeformed case, $Q_0$ and $Q_1$  correspond to the ground state and one particle state respectively.
Although not shown here, we have computed $Q_0$ and $Q_1$ to $L^4M^4$ order.
Then the energies of the ground state and one particle state are
\begin{align}
 \Delta_0^{(1)}=& \frac{\lambda  L^2 M^2}{8 \pi ^2 (J+1)}-\frac{\lambda L^4 M^4}{96 \pi ^2 (J+1)^2}+O\left(L^5M^5\right),  \\
 \Delta_1^{(1)}=& \frac{\lambda }{2 \pi ^2 \left(4 u_n^2+1\right)}+\frac{2 \lambda  L M u_n}{\pi
   ^2 (J+2) \left(4 u_n^2+1\right)}\nn\\
   & +\frac{\lambda  L^2 M^2 \left(4 (J (J
   (J+10)+20)+24) u_n^2+(J-6) (J+2)^2\right)}{8 \pi ^2 (J+2)^3 (J+3) \left(4
   u_n^2+1\right)}\nn\\
   &
-\lambda  L^3 M^3\frac{48 J^3 u_n^3+4 (J+2)^2 (J (4 J+7)+12) u_n}{3 \pi ^2 (J+2)^5 (J+3) (J+4) \left(4 u_n^2+1\right)}\nn\\
& -\lambda  L^4 M^4\frac{c_4 u_n^4+c_2 u_n^2+c_0 }{96 \pi ^2 (J+2)^7 (J+3)^3 (J+4) (J+5) \left(4 u_n^2+1\right)}+
   O\left(L^5M^5\right),
   \\
c_4=& 96 J^3 \left(J^5+36 J^4-92 J^3-1808 J^2-4656 J-2880\right),\nn\\
c_2=& 4 (J+2)^2 (J^8+18 J^7+495 J^6+2474 J^5+4520 J^4+12304 J^3
\nn\\&+35568 J^2+33696 J+17280),\nn\\
c_0=& (J+2)^4 (J+4) \left(J^5+6 J^4-211 J^3-738 J^2-756 J-1080\right). \nn
\end{align}
In large $J$ limit, we get
\begin{align}
 &\Delta_0^{(1)}=\frac{\lambda L^2  M^2}{8 \pi ^2 J}+
  (-\frac{\lambda L^2  M^2 }{8 \pi ^2 J^2}-\frac{\lambda L^4 M^4}{96 \pi ^2 J^2}+O\left(L^5M^5\right))+O(J^{-3}),  \\
  &\Delta_1^{(1)}=\frac{\lambda L^2  M^2}{8 \pi ^2 J}+
 (\frac{\lambda  n^2}{2 J^2}+\frac{\lambda L M n}{\pi  J^2}
 +\frac{\lambda L^2 M^2}{8 \pi ^2 J^2}
 -\frac{\lambda L^4 M^4 \left(\pi ^2 n^2+6\right)}{96 J^2 \left(\pi ^4 n^2\right)}
 +O\left(L^5M^5\right))+O(J^{-3}).
\end{align}
Assuming that the interaction between particles can be neglected in the large $J$ limit as in the undeformed case, the energy of an excited state above ground state energy is
\begin{equation}\label{ber}
  \Delta^{(1)}-\Delta_0^{(1)}=\sum_n N_n (\frac{\lambda n^2}{2 J^2}+\frac{\lambda L^2  M^2}{4 \pi ^2 J^2}-\frac{ \lambda L^4  M^4}{16 \left(\pi ^4 J^2 n^2\right)}+O\left(L^5M^5\right))+O(J^{-3}),
\end{equation}
where $N_n$ is excitation number for mode number $n$, and we assume that the total momentum is zero.

To compare the spectral curve result with the spin chain result, we expand $\Omega^{\hat 1 \hat 4}(x_n^{\hat 1 \hat 4})$ for small $LM$
and obtain the energy shift of the $sl(2)$ sector
\begin{equation}
  \d \D=\sum_{n} N_{n}^{\hat 1 \hat 4}\Omega^{\hat 1 \hat 4}(x^{\hat 1 \hat 4}_{n})
=\sum_n N_n^{\hat 1 \hat 4} (\frac{\lambda n^2}{2 J^2}+\frac{\lambda L^2  M^2}{4 \pi ^2 J^2}
-\frac{ \lambda L^4  M^4}{16 \left(\pi ^4 J^2 n^2\right)}+O\left(L^5M^5\right))+O(J^{-3}).
\end{equation}
where we have used  the level matching condition. The result agrees with (\ref{ber}).

The energy of the spin chain ground state to order  $L^6 M^6$ has been computed in \cite{Guica:2017mtd}
\begin{equation}
\begin{split}
\Delta_0^{(1)}&=
  \frac{\lambda }{4 \pi ^2 }
  (\frac{ L^2  M^2}{2(J+1)}-\frac{L^4  M^4}{24(J+1)^2 }+\frac{ L^6  M^6(J^2+J+2)}{720
  (J+1)^3(J+2) }+O\left(L^8M^8\right)),\\
 &=\frac{\lambda L^2 M^2}{8 \pi ^2 J }+ (-\frac{\lambda L^2  M^2}{8 \pi ^2 }-\frac{\lambda L^4  M^4}{96 \pi ^2 }+\frac{\lambda L^6  M^6}{2880 \pi ^2 }+O\left(L^8M^8\right))\frac{1}{J^2}+O\left(J^{-3}\right),
\end{split}
\end{equation}
The order $J^{-1}$ term matches the classical quantity $\Delta_{\mathrm{cl}}-J$, and the order $J^{-2}$ terms perfectly match
the one-loop shift $\Delta_{\mathrm{1-loop}}$ given in (\ref{1ls}).

\section{Conclusion and discussion}

In this paper we study the algebraic curve for superstring in $Sch_5 \times S^5$ and its application to the spectral problem. The asymptotic properties of the quasi-momenta for strings in $Sch_5 \times S^5$ are  nontrivial.
The point at infinity is a branch point of a cut connecting two Riemann sheets. We compute the semiclassical spectrum of the BMN string.
Remarkably, we show that in the large $J$ limit the string results match the gauge field results obtained by Baxter equation.
We provide a detailed  test of the Schr\"odinger/dipole CFT duality.

Our results  encourage  further exploration of integrability in  Schr\"odinger/dipole CFT duality.
 It would be nice to derive  the full quantum spectral curve of null dipole deformed $\mathcal{N} = 4$ super Yang-Mills theory, because the quasi-momenta are related to the quantum spectral curve in the strong coupling limit.
 It is also worth trying to obtain higher-order corrections on the field theory side to get a precise match with string theory predictions.
 One can also study the three dimensional  counterpart of  $Sch_5$, the warped $AdS_3$.
 We hope that integrability  would be  a powerful tool for the spectral problem of warped $AdS_3$/dipole CFT duality\cite{Song:2011sr}.

\section*{Acknowledgments}
The author would like to thank Nikolay Bobev, Hui-Huang Chen, Fedor Levkovich-Maslyuk, Zhibin Li, Wei Song, Jun-Bao Wu and Konstantin Zarembo for very helpful discussions and comments.
This work is supported in part by the National Natural Science Foundation of China (Grant No. 11575202).

%



\providecommand{\href}[2]{#2}\begingroup\raggedright\endgroup

\end{document}